\documentclass[11pt]{article}
\pdfoutput=1

\newcommand{\textSpacing}{1}
\newcommand{\algorithmSpacing}{1}
\renewcommand{\baselinestretch}{\textSpacing} 
\newcommand{\myedit}[2]{#2}

\usepackage{latexsym}
\usepackage{graphicx}
\usepackage{amsmath}
\usepackage{natbib}
\usepackage{multirow}
\usepackage{algorithm}
\usepackage{algorithmic}
\usepackage{listings}
\usepackage{url}

\def\T{^\prime}

\newcommand{\prob}[1]{\ensuremath{\text{Pr}\left(#1 \right)}}

	\newcommand{\bs}[1]{\boldsymbol{#1}}

\newcommand{\numDrugs}{J}
\newcommand{\logLike}{L}

\newcommand{\designmatrix}{{\bf X}}
\newcommand{\alldata}{{\bf Y}}
\newcommand{\alloffs}{{\bf L}}
\newcommand{\allloads}{{\bf M}}
\newcommand{\allnums}{{\bf N}}
\newcommand{\columndim}{K}
\newcommand{\vecOp}[1]{\mbox{vec}\left( #1 \right) }
\newcommand{\drug}{j}
\newcommand{\mess}{{\bf W}}
\newcommand{\deltaBeta}{\Delta \beta}

\newcommand{\gradientLogLike}{g_{\drug}(\bs{\beta})}
\newcommand{\hessianLogLike}{h_{\drug}(\bs{\beta})}

\newcommand{\prior}[1]{p\left( #1 \right)}
\newcommand{\XBeta}{\designmatrix\bs{\beta}}
\newcommand{\order}[1]{{\cal O}\left( #1 \right)}
\newcommand{\maxColumnSize}{X_{\mbox{\tiny max}}}

\newcommand{\iteration}{t}
\newcommand{\numerator}{\allloads \left[ \alloffs \times \exp(\XBeta) \times \designmatrix_{\drug} \right]}
\newcommand{\denominator}{\allloads \left[ \alloffs \times \exp(\XBeta) \right]}
\newcommand{\X}{{\bf X}}

\newcommand{\numCores}{C}
\newcommand{\partialSum}{\mbox{\sf PARTIAL\_SUM}}
\newcommand{\blockSize}{\mbox{\sf BLOCK\_SIZE}}
\newcommand{\numNonZero}[1]{\# \left( #1 \right)}
\newcommand{\betamap}{\hat{\bs{\beta}}_{\mbox{\tiny MAP}}}
\newcommand{\betamapj}[1]{\hat{\beta}_{ #1} }
\newcommand{\smalldrug}[1]{\mbox{\tiny #1}}

\begin{document}

\bibliographystyle{sysbio}


\title{Massive parallelization of serial inference algorithms for a complex generalized linear model}

\author{Marc A.~Suchard$^{1,2,3}$, Shawn E.~Simpson$^{4}$, Ivan Zorych$^{4}$, Patrick Ryan$^{5}$, David Madigan$^{4}$}

\date{}

\maketitle
 
\vspace{-1.2cm}
\begin{center}
  $^1$Department of Biomathematics and  
  $^2$Department of Human Genetics, \\
  David Geffen School of Medicine at UCLA, and \\
  $^3$Department of Biostatistics, UCLA School of Public Health, \\ 
  University of California, Los Angeles, CA, USA \\
  $^4$Department of Statistics, \\ Columbia University, New York, NY, USA \\
  $^5$Johnson \& Johnson Pharmaceutical Research and Development \\ Titusville, NJ, USA \\ 
\end{center}


\begin{abstract}
Following a series of high-profile drug safety disasters in recent
years, many countries are redoubling their efforts to ensure the
safety of licensed medical products.  
Large-scale observational databases such as claims databases or electronic health record systems are attracting particular attention in this regard, but present significant methodological and computational concerns. 
In this paper we show how high-performance statistical computation, including
graphics processing units, relatively inexpensive highly
parallel computing devices, can enable complex methods in large
databases.
We focus on optimization and massive parallelization of cyclic coordinate descent approaches to fit a conditioned generalized linear model involving tens of millions of observations and thousands of predictors in a Bayesian context.
We find orders-of-magnitude improvement in overall run-time.
Coordinate descent approaches are ubiquitous in high-dimensional statistics and the algorithms we propose open up exciting new methodological possibilities with the potential to significantly improve drug safety.
\end{abstract}


\section{Motivation and Background}

Increasing scientific, regulatory and public scrutiny focuses on the
obligation of the medical community, pharmaceutical industry and
health authorities to ensure that marketed drugs have acceptable
benefit-risk profiles \citep{coplan2010development}.  
%
Longitudinal observational databases provide
time-stamped patient-level medical information, such as periods of
drug exposure and dates of diagnoses, and are emerging as important data sources for associating the occurrence of adverse events (AEs) with specific drug use in the post-marketing setting once drugs are in wide-spread clinical use \citep{stang2010advancing}.
Some relevant papers focusing on drug
safety from observation databases include 
\citet{curtis2008adaptation,jin2008mining,li2009conditional,noren2008temporal,schneeweiss2009high,kulldorff2011maximized} 
Typical examples of these observation databases encompass administrative medical
claims databases and electronic health record systems, with larger claims databases containing upwards of 50 million lives with up to 10 years of data per life and exposure to 1000s of different drugs \citep{ryan2011learning}.
\par
The scale of these massive databases presents compelling computational challenges when attempting to
%
%
%
%
%
%
%
%
%
estimate the strength of association between each drug and each of several
AEs, while appropriately accounting for covariates such as other concomitant
drugs, patient demographics and concurrent disease.
Generalized linear models (GLMs) with unknown parameter regularization or Bayesian priors offer one thriving opportunity to estimate association strength while controlling for many other covariates \citep{madigan2011bayesian}.
However, naive implementation even to find maximum \textit{a posteriori} (MAP) point-estimates in standard statistical packages
grind to an almost stand-still with millions of outcomes and thousands of predictors, and hope of estimating even poor measures of uncertainty on drug-specific association estimates vanishes. 
\par
One usual approach to the computationally intensive task of statistical model fitting entertains distributing the work across a specialized and costly cluster or cloud of computers.
This approach achieves most success when the distributed jobs consist of lengthy ``embarrassingly parallel'' (EP) operations, such as the independent simulation of whole Markov chains in MCMC \citep{wilkinson2006parallel}.
However, the cluster with its distributed memory commands high
communication latency between operations, rendering even MAP estimation in a
GLM often unworkable, let alone estimation of second order terms such
as standard errors.  MAP estimation is generally an iterative algorithm,
and the potentially parallizable work within each step is rarely
sufficient to outweigh the communication latency and thread
management costs.
%
%
\par
For massive parallelization that overcomes many of these issues,
there exists a much less expensive resource available in many desktop and laptop computers,
the \textit{graphics processing unit} (GPU); see, for example, \citet{suchard2010understanding} for a gentle introduction in statistical model fitting.
GPUs are dedicated numerical processors originally designed for rendering 3-dimensional computer graphics.
A GPU consists of tens to hundreds of processor cores on a single chip. These can be programmed to perform a
sequence of numerical operations simultaneously to each element of a large data array. The acronym SPMD
summarizes this single program, multiple data paradigm.  
Because the same operations, called kernels,
function simultaneously, GPUs can achieve extremely high arithmetic intensity provided one can transfer
input data and output results onto and off of the processors efficiently.  Because the parallel threads driving the kernels operate on the same computer card, the cost of spawning and
destroying millions of threads within each iterative step of the MAP estimation is neglible, and communication latency 
between threads is minimal.  
However, statisticians have been slow to embrace the new technology, due in part to a preconception that GPUs work best with EP operations.
To dispell these ideas, \citet{silberstein2008efficient} first demonstrated the potential for GPUs in fitting simple Bayesian networks.  Recently, \citet{suchard2009many} and \citet{suchard2010understanding} have seen greater than $100$-fold speed-ups in MCMC simulations involving highly structured graphical models and mixture models.
\citet{lee2010utility} and \citet{tibbits2009parallel} are following suit with Bayesian model fitting via particle filtering and slice sampling, and \citet{zhou2010graphics} demonstrate GPU utility for high-dimensional optimization. 
\par
In this paper, 
we explore the use of GPU parallelization in fitting a real-world problem involving a penalized GLM to massive observation datasets with high-throughput computing needs.
We entertain recent advances in a Bayesian self-controlled case series (BSCCS) model \citep{madigan2011bayesian} 
and by
exploiting the sparsity of the resulting database design matrix, we optimize a cyclic coordinate descent (CCD) algorithm to generate MAP estimates for this high-dimensional GLM.
Given the substantial speed-up that optimization and the GPU afford, we provide for the first time rough estimates of the prior hyperparameters 
and limited measures of coefficient uncertainty.

\section{Methods}

\myedit{NewGLMIntro}{
GLMs assume subject outcomes arise from an exponential family distribution whose mean is a deterministic function of an outcome-specific linear predictor \citep{nelder1972generalized}.  These models include, for example, logistic regression,  Poisson regression and several survival models.  Often, study designs necessiate matching subjects or conditioning on sufficient statistics of the generative distribution to infer the relative effects of predictors; this leads to complex GLMs with likelihood-functions that grow computationally expensive in massive datasets.  We explore one such model as a case-study in optimization and parallelization.
}
\subsection{Bayesian Self-Controlled Case Series Model}

\citet{farrington1995relative} proposed the \emph{self-controlled case series}
(SCCS)
method in order to estimate the relative incidence of AEs
to assess vaccine safety. 
SCCS
provides a cases-only (subjects with at least one AE) design that automatically
controls for time-fixed covariates.
Since each subject serves as her own control, all individual-specific effects 
drop out of the model likelihood and the method compares AE rates between exposed and unexposed
time-intervals through an underlying inhomogeneous Poisson process assumption.
%
%
%
%

\begin{figure}
\centerline{\includegraphics[width=0.75\textwidth]{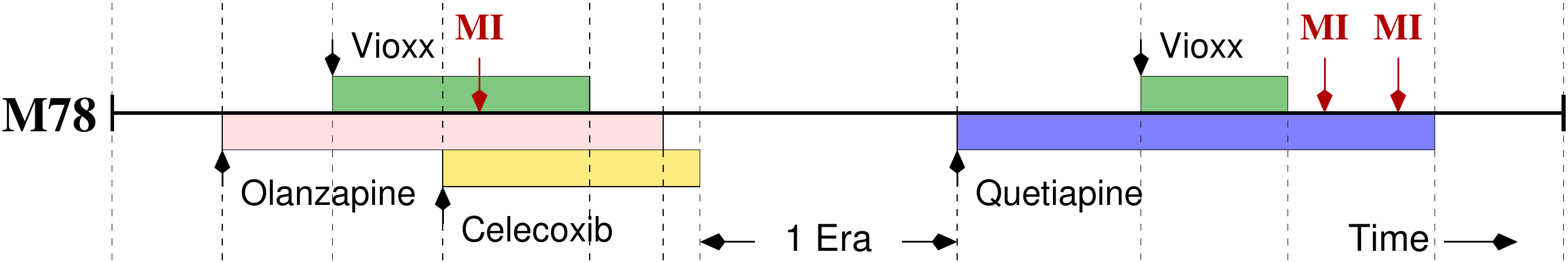}}
\caption{Representative drug exposure and adverse event (myocardial infarction, MI) history for one 78 year-old male.  In constant exposure era $k = 4$, this subject suffers an MI ($y_{i,4} = 1$) and is taking Vioxx, Olanzapine and Celecoxib ($x_{i,4,\smalldrug{V}} = x_{1,4,\smalldrug{O}} = x_{i,4,\smalldrug{C}} = 1$).  In era $k = 10$, this subject suffers two MIs ($y_{i,10} = 2$) and is taking Quetiapine ($x_{i,10,\smalldrug{Q}} = 1$).
\label{history-figure}
}
\end{figure}

Suppose an observational database tracks the drug exposure and AE history of $i = 1,\ldots,N$ subjects who each experience at least one AE.
Figure \ref{history-figure} cartoons such a history.
During observation, subjects start and stop individual regiments of $\drug = 1,\ldots, \numDrugs$ possible drugs accumulated across all subjects.  These regiments partition each subject's observational period into $k = 1, \ldots, K_i$ eras, during which drug exposure is (assumed to be) constant.
Drug exposure indicators ${\bf x}_{ik} = (x_{ik1},\ldots,x_{ik\numDrugs})\T$ and time-lengths $l_{ik}$ (generally recorded in days) fully characterize each era for each subject, where $x_{ik\drug} = 1$ if exposed to drug $\drug$ or otherwise 0.
Finally, $y_{ik}$ counts the number of AEs for subject $i$ during era $k$ and, for convenience, we group ${\bf y}_i = (y_{i1}, \ldots, y_{i{K_i}} )\T$, ${\bf x}_i = ( {\bf x}_{i1}, \ldots, {\bf x}_{i{K_i}} )\T$ and ${\bf l}_i = (l_{i1}, \ldots, l_{i{K_i}} )\T$. 

\newcommand{\placeHolder}[1]{
\par
\centerline{[#1 about here.]}
\par
}

%
%
%
%
%
	
A SCCS assumes that these AEs arise according to an inhomogeneous Poisson process, where a subject baseline effect $e^{\phi_i}$ and drug exposures multiplicatively modulate the underlying instantaneous event intensity
$
	\lambda_{ik} = e^{ \phi_i + {\bf x}_{ik}\T \bs{\beta} }
$	
for subject $i$ during era $k$. Here, $\bs{\beta} = (\beta_{1},\ldots,\beta_{\numDrugs})\T$ are unknown relative risks attributable to each drug. Consequentially, $y_{ik}\ \sim\ \mbox{Poisson}(\lambda_{ik})$.  
Due to computational demand, researchers have typically applied this model to study only one potential exposure at a time, ignoring correlation between drugs.
\citet{madigan2011bayesian}
provide further details on the development of the multivariate SCCS involving multiple drugs simultaneously as used here.
\par
In order to avoid estimating parameters $\phi_i$ for all $i$, the SCCS method conditions on their sufficient statistics.
Under the Poisson assumptions, these sufficient statistics are the total number of AEs $n_i = \sum_k y_{ik}$ that a subject experiences over her observation period, yielding the model likelihood contribution for each subject,
\begin{equation}		
			\prob{ {\bf y}_i \mid {\bf x}_i, \, n_i }
			= \frac{ \prob{ {\bf y}_i \mid {\bf x}_i } }{ \prob{ n_i \mid {\bf x_i} } }
			\propto \prod_{k=1}^{K_i} 
							\left( 
								\frac{ e^{\, {\bf x}_{ik}\T \bs{\beta}} }{ \sum_{k'} l_{ik'} e^{\, {\bf x}_{ik'}\T \bs{\beta} } } 
							\right)^{\!\! y_{ik}} .
\end{equation}
%
%
Naturally, it is also clear from the likelihood expression that if a subject experiences no AEs ($n_i = 0$), then
that subject does not contribute to the model likelihood, providing a cases-only design.  
\par
Taking all subjects as independent, we write the log-likelihood as a function of unknown $\bs{\beta}$ as
\begin{equation}
		\logLike\left( \bs{\beta} \right) = \sum_{i=1}^N \left[ 
				\sum_{k=1}^{K_i} \left( y_{ik} \, {\bf x}_{ik}\T \bs{\beta} \right)
				- n_i \, \log \left( \sum_{k=1}^{K_i} 
				l_{ik} \, e^{{\bf x}_{ik}\T \bs{\beta}} 
				\right)
			\right]	
			.
\label{logLikeUgly}
\end{equation}
This likelihood furnishes
a complex GLM that carries a high computational burden, arising from the conditioning and renormalization.
\myedit{Burden}{
In practice, one needs to keep track of the sum of a large number of terms for each subject and each term requires exponentiation and then weighting.
}
Such a burden quickly grows prohibitive for the millions of cases available in observational databases.

\paragraph{Priors}

In drug safety surveillance there exist thousands of potential drugs.
This high dimensionality can lead to severe overfitting under the usual maximum likelihood approach, even for massive datasets, so regularization remains necessary.
As an alternative, we adopt a Bayesian approach by assuming a prior $\prior{\bs{\beta}}$ over the drug effect parameter vector, constructing a 
BSCCS
\citep{madigan2011bayesian}
and performing inference based on posterior mode estimates.
\myedit{Penalize}{
We refer interested readers to \cite{kyung2010penalized} for a more in-depth discussion of the connections between penalized regression and some Bayesian models.
}
\par
To develop $\prior{\bs{\beta}}$, we naturally assume that most drugs have no appreciable effect and consider distributions that shrink the parameter estimates toward or to $0$ to also address overfitting.  We focus on two choices:
\begin{equation}
	\beta_{\drug} \sim \mbox{Normal}(0, \sigma^2) \quad \mbox{ or } \quad \beta_{\drug} \sim \mbox{Laplace}(0, \sigma^2)
\end{equation}
for all $j$, where $\sigma^2$ is the unknown hyperparameter variance of each distribution.
Under the Normal prior, finding the posterior mode estimates is analogous to ridge conditioned-Poisson regression with its $L_2$-norm constraint on $\bs{\beta}$, and, under the Laplacian prior, we achieve a lasso conditioned-Poisson regression with its $L_1$-norm constraint \citep{tibs1996lasso}.

\subsection{Maximum \textit{A Posteriori} Estimation using Cyclic Coordinate Descent}

CCD algorithms \citep{d1959convex,warga1963minimizing} for fitting generalized linear models with $L_1$ or $L_2$ regularization priors come in many flavors \citep{wu2008coordinate}.
The overarching theme of these algorithms promotes forming a fixed or random cycle over the regression parameters $\bs{\beta}$ and updating one element $\beta_{\drug}$ at a time, achieving after iteration their maximum \textit{a posteriori} estimates $\betamap = (\betamapj{1},\ldots,\betamapj{\numDrugs})$.  
These updates require evaluating the log posterior gradient
$\frac{\partial}{\partial \beta_{\drug}} P (\bs{\beta})$
and Hessian $\frac{\partial^2}{\partial \beta_{\drug}^2} P (\bs{\beta})$, where $P(\bs{\beta}) = L(\bs{\beta}) + \log \prior{\bs{\beta}}$, along a single dimension only, and thus avoid the ``Achilles heel'' \citep{wu2009genome} of the more standard multivariate Newton's method that necessiates inverting the complete and high-dimensional Hessian at each iteration.
\par
Within the cycle, CCD implementations often differ in the size of the one-dimensional step $\deltaBeta_{\drug}$ they take.  
The traditional algorithm proposes iterating one-dimensional Newton's method updates to convergence.  
Others consider a single-step update based (sometimes loosely) on Newton's method, where one bounds the second derivates or $\deltaBeta_{\drug}$ directly to ensure a descent property and minimize algebraic work \citep{lange1995gradient,dennis1989chapter,zhang2001text,genkin2007large,wu2008coordinate}.
These single-step algorithms often escape the excess overhead of monitoring for convergence of the single-parameter Newton's methods.  
\par
To explore MAP estimation via CCD for the BSSCS model, we follow the success of \cite{genkin2007large} and employ an adaptable trust-region bound on $\deltaBeta_{\drug}$, where the unbounded $\deltaBeta_{\drug}$ follows from a single application of Newton's method \citep{zhang2001text}:
\begin{equation}
\deltaBeta_{\drug} =  - \frac{
\frac{\partial}{\partial \beta_{\drug}} \left[  
\logLike (\bs{\beta}) + \log \prior{\bs{\beta}}
\right]
}{
\frac{\partial^2}{\partial \beta_{\drug}^2} \left[  
\logLike (\bs{\beta}) + \log \prior{\bs{\beta}}
\right]
}
=
- \frac{
\gradientLogLike + 
\frac{\partial}{\partial \beta_{\drug}} 
\log \prior{\bs{\beta}}
}{
\hessianLogLike + 
\frac{\partial^2}{\partial \beta_{\drug}^2} 
\log \prior{\bs{\beta}}
}
.
\end{equation}
\myedit{TrustRegion}{

}
We outline the complete fitting procedure in Algorithm \ref{ccd-algorithm}.
Following \cite{genkin2007large}, we declare convergence when the sum of the absolute change in $\XBeta$ from successive iterations falls below $\epsilon = 0.0005$.  
\myedit{CurrentAdvances}{
	The preceeding approach has been effective in fitting the BSSCS model to modest datasets \citep{simpson2011self}.  Our present inquire in what follows attempts to extend this success to massive observation databases.
}

\renewcommand{\baselinestretch}{\algorithmSpacing} 
\begin{algorithm}
\begin{algorithmic}[1]
\medskip
\STATE Initialize: $\bs{\beta} = \bs{0}$ which implies 
$\left[ \designmatrix{\bs{\beta}} \right] = \bs{0}$,
$\left[ \alloffs \times \exp \left( \designmatrix\bs{\beta} \right) \right] = \alloffs$, and
$ \allloads \left[ \alloffs \times \exp \left( \designmatrix\bs{\beta} \right) \right] = \allloads \alloffs$
\STATE Initialize: outer iteration counter $\iteration = 1$
\REPEAT     
	\FOR{$\mbox{inner cycle}\ \drug=1$ to $\numDrugs$}
		\STATE Compute unidirectional gradient $\gradientLogLike$ and Hessian $\hessianLogLike$ (target for parallelization)
		\STATE Compute $\deltaBeta_{\drug}$ given $\gradientLogLike$, $\hessianLogLike$ and derivatives of prior $\prior{\bs{\beta}}$
		\IF{$\deltaBeta_{\drug} \neq 0$}
			\STATE $\bs{\beta} \leftarrow \bs{\beta} + \left( \deltaBeta_{\drug} \right) {\bf e}_{\drug}$
			\STATE Update 
			$\left[ \designmatrix\bs{\beta} \right]$, 
			$\left[ \alloffs \times \exp{\designmatrix\bs{\beta}} \right]$ and
			\allloads $\left[ \alloffs \times \exp{\designmatrix\bs{\beta}} \right]$
			(target for parallelization)
		\ENDIF
	\ENDFOR
	\STATE 
	Update $\iteration \leftarrow \iteration + 1$
\UNTIL{convergence in $\designmatrix\bs{\beta}$ occurs}
\STATE Report: $\betamap$ and maximized log-posterior $P ( \betamap )$
\end{algorithmic}
\caption{Cyclic coordinate descent algorithm for fitting Bayesian self-controlled case series model.  Computationally demanding steps are highlighted as targets for parallelization.  While all variables are defined in the text, we identify $\bs{\beta}$ here as the $\numDrugs$-dimensional regression coefficients over which we wish to maximize the log-posterior in this algorithm.\medskip
}
\label{ccd-algorithm}
\end{algorithm}
\renewcommand{\baselinestretch}{\textSpacing} 

\lstset{numbers=left, numberstyle=\tiny, basicstyle=\small, language={C++}, 
morekeywords={real,__global__}
}


\subsection{Computational Work}

To gain a handle on the computational work involved in fitting the BSCCS model, 
let $\columndim = \sum_i^N K_i$ count the total number of (unique within subject) exposure eras 
across all subjects.  Define 
$\designmatrix = \vecOp{ 
{\bf x}_{ik}\T 
}$ as the sparse $\columndim \times \numDrugs$ design matrix that consists solely of $0$ and $1$ entries, indicating if drug $\drug$ contributes to each of the $\columndim$ patient/exposure-era rows.  Likewise, form $\alldata = (y_{11},\ldots,y_{N K_N})\T$ and $\alloffs = (l_{11},\ldots,l_{N K_N})\T$
as $\columndim$-dimensional column vectors, $\allnums = (n_1,\ldots,n_N)\T$ as an $N$-dimensional column vector and $\allloads$ as a sparse $N \times \columndim$ loading matrix with entries
\begin{equation}
    M_{ik}
    = \left\{
	\begin{array}{l}
	1 \mbox{ for }  
    \sum_{m = 1}^{i - 1} K_m < k \le \sum_{m=1}^{i} K_m \\
	0 \mbox{ otherwise.}
	\end{array}
	\right.
\end{equation}

Making these substitutions into Equation (\ref{logLikeUgly}), we achieve
\begin{equation}
\logLike \left( \bs{\beta} \right) = 
	\alldata\T\designmatrix\bs{\beta} - \allnums\T 
	\log \left\{ \allloads \left[   
		\alloffs \times \exp
		\left(
			\designmatrix\bs{\beta}
		\right) 	
	\right] \right\},
\end{equation}
where we have defined multiplication $(\times)$, exponentiation $(\exp)$ and forming the logarithm $(\log)$ of a column-vector as element-wise operations.
It remains possible to avoid the Hadamard product definition of element-wise multiplication and, to come shortly, division $(/)$ in favor of standard matrix-multiplication and matrix-inversion by exploiting a reparameterization of the loading matrix and diagonal matrices.  However, their use belies the simplicity of the element-wise, and hence highly parallelizable, operations we encounter in practice in computing the unidimensional gradients and Hessians.
Differentiating $\logLike$ with respect to $\beta_{\drug}$ returns the necessary unidimensional gradient 
\begin{align}
	\gradientLogLike &= \frac{\partial{\logLike}}{\partial{\beta_{\drug}}} = 
	\alldata\T \designmatrix_{\drug} - \allnums\T \mess, 
\end{align}
where
\begin{align}
	\mess &=
\frac{
	\allloads \left[ 
	\alloffs \times	
	\exp\left(\designmatrix\bs{\beta}\right) \times \designmatrix_{\drug}
	\right]
}{
	\allloads \left[
	\alloffs  \times \exp\left(
			\designmatrix\bs{\beta}
	\right)	
	\right]
}		
	,
	\label{compute-gradient}
\end{align}
and vector $\designmatrix_{\drug}$ is the $\drug^{\mbox{\tiny th}}$ column of $\designmatrix$.
%
Likewise, the relevant entry of the Hessian matrix falls out as
\begin{equation}
 	\hessianLogLike = \frac{\partial^2{L}}{ \partial{\beta_{\drug}^2}} =
- \allnums\T 
\left[
	\mess \times
	\left(
	{\bf 1} - \mess	
	\right)
\right]
	.
 	\label{compute-hessian}
\end{equation}


\subsection{Targets for Parallelization}

CCD, along with most forms of statistical optimization and Markov chain Monte Carlo, is an inherently serial algorithm. 
As reminded in Algorithm \ref{ccd-algorithm}, even within a iteration $\iteration$, one cycles over parameters $\drug$ to update and work cannot begin computing the next parameter update until the current update completes.
Such algorithms do not immediately appear amenable to parallelization.
However, all is not lost when one considers the proportion of computational work performed within each update to the computational overhead of the serial component.
CCD carries a surprisingly light-weight serial component, and for the BSCCS model applied to even the smallest observational database described below, over 99.5\% of the run-time lies in computing $\gradientLogLike$ and $\hessianLogLike$ alone.
\par
To provide insight for readers who wish to explore massive parallelization in their own applications, we study the computational complexity of evaluating $\gradientLogLike$ and $\hessianLogLike$ and, in the process, identify likely targets for optimization and parallelization.
Common to both $\gradientLogLike$ and $\hessianLogLike$ is $\mess$; hence efficient computation proceeds via first evaluating $\numerator$ and $\denominator$ that comprise $\mess$. To compute these, we
\begin{enumerate}
	\item Update $\left[ \XBeta \right]$ -- given $\XBeta$ and $\deltaBeta_{\drug - 1}$ from the previous iteration, $\XBeta \leftarrow \XBeta + \deltaBeta_{\drug - 1} \designmatrix_{\drug - 1}$.  When $\X$ is dense, the serial complexity of this operation is $\order{\columndim}$.  For sparse $\X$,
	the worst-case complexity decreases to $\order{\maxColumnSize}$ where $\maxColumnSize$ is the maximum of $\numNonZero{\X_{\drug}}$ over $\drug = 1,\ldots,\numDrugs$ and $\numNonZero{\X_{\drug}}$  counts the number of non-zero entries in $\designmatrix_{\drug}$.  In general, $\maxColumnSize \ll \columndim$. \label{xbeta-step}
	
	\item Evaluate or update $\left[ \alloffs \times \exp(\XBeta) \right]$ -- while this is also a $\columndim$-dimensional vector, only the elements for which $\X_{\drug-1}$ are non-zero have changed; therefore, this step either re-evaluates all elements with $\order{\columndim}$ or updates a few elements in $\order{\maxColumnSize}$.  In both cases,
	the scaling constant is large because computing $\exp(x)$ requires 10s to 100s of times longer than elementry floating point operations. \label{exp-step} 
	
	\item Evaluate or update $\allloads \left[ \alloffs \times \exp(\XBeta) \right]$ -- for dense $\X$, this a sparse-matrix/dense-vector multiplication; $\numNonZero{\allloads} = \columndim$, achieving $\order{\columndim}$.  
	When $\X$ is sparse, it remains faster to update just the affected elements in $\order{\maxColumnSize}$; see Listing \ref{sparse-kernel} for details of this update. \label{denominator-step}
	
	\item Evaluate $\allloads \left[ \alloffs \times \exp(\XBeta) \times \designmatrix_{\drug} \right]$ -- here we find either a sparse-matrix/dense-vector multiplication or an unusual sparse-matrix/sparse-vector multiplication with worst-case complexity $\order{\maxColumnSize}$ 
	\label{numerator-step}
\end{enumerate}
Steps (\ref{xbeta-step}) through (\ref{denominator-step}) depend on $\X_{\drug -1}$ and convenience suggests performing these steps at the end of the previous iteration to reduce book-keeping.
We illustrate this point in Algorithm \ref{ccd-algorithm}.
Also noted in Algorithm \ref{ccd-algorithm} is the observation that these steps only need envoking when $\deltaBeta_{\drug} \neq 0$.
For the Laplacian prior with its discontinuity at $0$, $\deltaBeta_{\drug} = 0$ occurs regularly.

\paragraph{Exploiting Sparsity}

Starting with Step (\ref{xbeta-step}) above, a very naive implementation recomputes the matrix-vector product at each cycle with complexity $\order{\columndim \numDrugs}$ and the potential to drive run-time to a dead-lock.
\cite{zhang2001text} and \cite{wu2009genome} independently identify the savings that the one-dimensional update affords here.
These works, along with \cite{genkin2007large}, exploit the sparsity of $\X$ in updating the dense $\columndim$-dimensional vector $\XBeta$ (Step \ref{xbeta-step}).  For comparison, these papers refer to $\XBeta$ as $(r_1,\ldots,r_N)$.
However, we are unaware of others who continue to exploit the sparsity of $\X$ in moving from $\XBeta$ through to the subject-specific components of the gradient and Hessian (Steps \ref{exp-step} - \ref{numerator-step}).
%
\par
With the numerator and denominator components of $\mess$ in hand, we form a simple element-wise transformation and take two simultaneous inner products to return $\allnums\T \mess$ and $\allnums\T \left[ \mess \times \left(\bs{1} - \mess\right)\right]$.
We discuss the advantages of these fused reductions for both host CPUs and GPUs shortly.
This operation carries worst-case serial complexity $\order{N}$ when all subjects have at least one exposure era that the current drug influences.  Since several drugs carry prevalences among subjects nearing 25 - 50\%, keeping track of the non-zero elements in $\mess$ and performing sparse operations often reduces efficiency given the extra over-head and irregular memory access.  In either case, the work in this fused-reduction is far greater than the work required to compute the one-dimensional gradient and Hessian contributions of the prior on $\beta_\drug$, so we leave the prior details to the reader.
Finally, after cycling over all drugs $\drug$, we evaluate the convergence criterion.
Whether one checks the change in $\XBeta$ \citep{genkin2007large} or in the log-posterior, these computations remain a daunting $\order{\columndim}$.  Fortunately, we only envoke them once per complete cycle and this task's run-time becomes nearly irrelevant for moderate $\numDrugs$.
\par
\paragraph{Fine-Scale Parallelization}
From these computational complexities, we immediately identify that Step (\ref{exp-step}) dominates run-time when $\X$ is dense at $\order{\columndim}$ and with a very large scalar constant.
On the other hand, for sparse $\X$, the fused reduction at $\order{N}$ trumps run-time.
Fortunately, these operations, along with the other update steps, are prime targets for parallelization using GPUs.
\par
\renewcommand{\baselinestretch}{\algorithmSpacing} 
\begin{lstlisting}[frame=tb,texcl,float,caption={Dense CUDA kernel for element-wise evaluation of 
$\alloffs\times \exp( \XBeta)$ given $\alloffs$ and $\XBeta$
\label{dense-kernel}
}]
__global__ void evaluateLExpXBeta(
		real* LExpXBeta, const int* L,
		const real* XBeta, int K) {		
	// Determine element index for this thread
	int idx = blockIdx.x * blockDim.x + threadIdx.x;
	// Perform scalar operation on each vector element
	if (idx < K) {
		// All coalesced memory access
		LExpXBeta[idx] = L[idx] * exp(XBeta[idx]);
	}								           
}
\end{lstlisting}
\renewcommand{\baselinestretch}{\textSpacing}
Code Listing \ref{dense-kernel} presents a basic CUDA kernel to perform the dense computation of Step (\ref{exp-step}).  
To envoke this kernel, the host program requests the short-lived execution of $\columndim$ threads, one for each computed element.
Unlike coarser-scale parallelization using MPI across clusters or even multi-core approaches, the cost of creating and destroying threads is often negligible for GPUs; this makes such fine-scale parallelization ideal for embedding within serial algorithms.
While each thread executes independently in this kernel as there is no shared data between threads, we still group threads into relatively large thread-blocks of size, say, $8 \times 16$ or $16 \times 16$.
For all NVIDIA hardware, 16 sequential global memory read/writes ``coalesce'' into a single transaction, and the GPU interleaves the execution of multiple 16-thread sets 
in the same block to hide transaction latency.  
While recent GPUs relax the sequential requirement modestly, both processes significantly decrease memory-bandwidth limitations, improving arithmetic throughput. 
All of the work of this kernel falls in a single line of code.  The theoretical complexity of this operation in parallel reduces to $\order{1}$; however, in practice, one achieves $\order{\columndim / \numCores}$ where $\numCores$ counts the number of GPU processing cores available to the host.  This quantity can range from the low-10s on an integrated GPU in a mobile device or laptop to the mid-100s on a dedicated GPU card in a desktop through to the low-1000s on multiple GPU devices attached to a single host.
\par
\paragraph{Fused Operations}
We turn our attention to the element-wise transformation and simultaneous inner products that provide a noteworthy example of effective optimization for massive datasets.
Fusing these steps 
into a single operation avoids explicitly forming $\mess$, writing to its location in memory $N$ times and then immediately reading from it $N$ or $2N$ times, depending on if we perform the inner products simultaneously or separately, for each cycle step.  
Further, we only need to read from $\allnums$ once during the simultaneous reduction.
This can significantly reduce memory bandwidth requirements on both the host CPU and GPU.
Memory bandwidth measures the rate at which the processor can read data from or store data to memory.
With increasingly faster processor speeds, many algorithms in statistics are memory bandwidth-limited rather than arthimetic throughput-limited.
These optimization strategies fall under the name of ``lazy evaluation'' to ``reduce temporaries'' and warrant greater recognition among computational statisticians who provide high-performance tools.
Modern computing language compilers are very proficient at optimizing away such intermediates for scalar operations, but often fail for vectors since large vectors cannot be held entirely in processor registers.
For CPU-computing, high-level linear algebra libraries, such as {\sf Eigen}, adeptly reduce the use of vector and matrix temporaries and provide lazy evaluation through ``expression templates'' \citep{veldhuizen1995expression}.
For the GPU, the {\sf Thrust} library furnishes expression templates for fusing a univariate-to-univariate transformation and reduction of a single input vector, and we highly recommend this tool.
However, currently, statisticians are hardpressed to find a GPU library that takes two input vectors, performs a bivariate-to-bivariate transformation and simultaneously reduces both output vectors.
While such functionality may initially appear convoluted, it finds use in efficiently computing the one-dimensional gradient and Hessian for any GLM with minor modification to the transformation.
To this end, we hand-craft our own.
\par
\renewcommand{\baselinestretch}{\algorithmSpacing} 
\begin{lstlisting}[frame=tb,texcl,float,caption={Fused CUDA kernel for 
transformation and reduction of numerators $\numerator$, denominators $\denominator$ and $\allnums$ to partial-sums of $\gradientLogLike$ and $\hessianLogLike$.  Partial-sums end in length 
$\sf{PARTIAL\_SUM}$
and we further reduce these on the host for efficiency. Function 
$\sf{paralleReduction}$ 
performs a generic logarithimic-order reduction in shared memory.
\label{reduction-kernel}
}]
__global__ void fusedTransformationAndReduction(
        const real* Numerator, const real* Denominator, const int* N,
        real* Gradient, real* Hessian, int length) {
        
    // Define shared memory for thread-block reduction
    __shared__ real sGradient[PARTIAL_SUM], sHessian[PARTIAL_SUM];
        
    // Partial sums for this thread    
    real tSumGradient = 0.0, tSumHessian = 0.0;    
    
    // Determine first element index for this thread   
    int idx = blockIdx.x * PARTIAL_SUM + threadIdx.x;            
    while (idx < length) { // Each thread processes multiple entries
    
        // Do transform of this entry and add to local thread sum       
        real ratio = Numerator[idx] / Denominator[idx]; // Coalesced memory access
        int n = N[idx]; // Coalesced memory access
        real tGradient = n * ratio;        
        tSumGradient += tGradient;
        tSumHessian += tGradient * (1.0 - ratio);
        
        idx += PARTIAL_SUM * gridDim.x; // Index next element for thread
    }	
    
    // Reduce across all threads in block, leaves total in first element of shared memory
    parallelReduction(sGradient, tSumGradient);
    parallelReduction(sHessian, tSumHessian);	
    
    // Only one thread writes block result
    if (threadIdx.x == 0) {
        Gradient[blockIdx.x] = sGradient[0];
        Hessian[blockIdx.x]  = sHessian[0];
    }
}
\end{lstlisting}
\renewcommand{\baselinestretch}{\textSpacing}
Listing \ref{reduction-kernel} presents our fused CUDA kernel.  
To envoke this kernel, the host program requests the execution of a moderate number ($\partialSum \le 64$) of thread-blocks in which each block drives 256 or more ($\blockSize$) threads depending on the hardware.
In parallel, each thread begins by looping over $N / (\partialSum \times \blockSize)$ elements in $\numerator$, $\denominator$ and $\allnums$, forming their transform and accumulating both inner product contributions for these elements. 
We interleave which elements each thread visits to coalesce memory transactions.
Once completed, the threads within a block exploit the block's shared memory to perform two generic $\order{\log \blockSize}$ tree-based parallel reductions \citep{harris2010optimizing}.
While limited in quantity, shared memory on a GPU sports orders of magnitude faster access time than global memory and is accessible by all threads in the same block during their execution.  
Shared memory enables threads to conveniently share data, such as is required in the tree-reduction.
The CUDA software development kit (SDK) furnishes several examples of tree-based parallel reductions.
The output from this kernel are sets of partial-sums for $\gradientLogLike$ and $\hessianLogLike$, each of length $\partialSum \ll N$.
Instead of envoking a second round of parallel reduction on these partial-sums, we perform the final work in series on the host.
Because of high communication latency between the host and GPU device, it takes comparable time to transfer 2 floating-point values as the modest $2 \times \partialSum$.
CPU/GPU work-balance will be a hallmark for speed-efficient statistical fitting of massive datasets.

\par
In terms of work-balance, while the fused reduction is the rate limiting step for sparse $\X$ on both the CPU and GPU, we can significantly decrease the communication latency between the host and GPU by additionally off-loading all of Steps (\ref{xbeta-step}) through (\ref{numerator-step}) to the GPU well.  
Instead of uploading $2 \times N$ floating-point numbers to the GPU in each cycle step, we succeed in reducing this number to a single floating-point  $\deltaBeta_{\drug}$.
The cost, of course, is additional programming and the need for performing sparse operations on the GPU.
\par
\paragraph{Representation in Memory}
Memory access is often irregular for sparse linear algebra, and the computational statistician needs to pay particular attention to how both sparse matrices and vectors are represented in memory \citep{bell2009efficient,baskaran2009optimizing}.
For example, the optimal representations for $\designmatrix$ and $\allnums$ differ.
Only single columns of $\designmatrix$ enter into Steps (\ref{xbeta-step}) and (\ref{numerator-step}) at a time, highlighting the need for compressed column storage (CCS) in which one places consecutive non-zero elements of each column $\X_{\drug}$ into adjecent memory addresses.
While the standard representational choice for sparse-matrix/dense-vector (spMV) multiplication is compressed row storage (CRS), naive CRS representation of $\designmatrix$ would be detrimental to run-time on computing hardware with limited low-level caches, such as GPUs, since CRS is designed for row-by-row access.
On the other hand, loading matrix $\allloads$ does enter into $\mess$ as a spMV multiplication operation when $\X$ is dense, suggesting CRS.
For sparse $\X$, $\numNonZero{\X_{\drug}} \ll \numNonZero{\allloads}$, so precomputing $\allloads \X_{\drug}$ for all $\drug$ in coordinate (COO) representation is simple and effective.  
COO representation consists of two index arrays, one to hold the row-indicators and one to hold the column-indicators of the non-zero entries, held in a third value array.
Here the column-indicators of $\allloads \X_{\drug}$ conveniently are the same as the column-indicators of $\X_{\drug}$.
Stored as a structure of arrays (SoAs), memory access to the row- and column-indicators is sequential, well-cached on the host CPU and coalesced on the GPU.  However, retrieving the individual elements of $\alloffs\times \exp( \XBeta)$ remains irregular.  Finally, the non-zero entries of both $\designmatrix$ and $\allnums$ are all one, so they need not be stored.
\par
Executing an independent thread for each non-zero element of $\X_{\drug}$ to update $\numerator$ may result in race conditions when multiple threads attempt memory transactions on the same elements in $\numerator$ in global memory.
One solution entertains launching one or more cooperative threads tied to each of the $N$ rows in $\allloads \X_{\drug}$.
This may generate large inefficiencies as many rows in $\allloads \X_{\drug}$ contain only zeros.
Alternatively, the last few generations of GPUs contain small on-processor memory caches that enable relatively quick ``atomic'' transactions, in which only one thread may access a specific address in global memory at a time.
This avoids race conditions and allows us to fuse the sparse updates of Steps (\ref{xbeta-step}) through (\ref{denominator-step}) together into a single kernel.
%
%
%
%
Listing \ref{sparse-kernel} presents our sparse CUDA kernel to update $\XBeta$,
$\alloffs\times \exp( \XBeta)$, and
$\denominator$
given $\deltaBeta_{\drug}$.  
We envoke this kernel with one thread per non-zero entry in $\X_{\drug}$, grouping threads into large blocks to help hide memory latency.
\par
\renewcommand{\baselinestretch}{\algorithmSpacing} 
\begin{lstlisting}[frame=tb,texcl,float,caption={Sparse
CUDA kernel for updating
$\XBeta$,
$\alloffs\times \exp( \XBeta)$, and
$\denominator$
given $\X_{\drug}$ and $\deltaBeta_{\drug}$
\label{sparse-kernel}
}]
__global__ void sparseUpdate(
        real* XBeta, real* LExpXBeta, real* NLExpXBeta,
        const int* L, const int* sparse_rows, const int* sparse_columns,
        real DeltaBeta, int NumNonZeros) {
        
    // Determine sparse element index for this thread									
    int idx = blockIdx.x * blockDim.x + threadIdx.x;
    if (idx < NumNonZeros) {       
        int n = rows[idx];     // Coalesced memory access
        int k = columns[idx];  // Coalesced memory access        
        
        // Read sparse elements, many non-coalesced
        real oldXBeta = XBeta[k];
        real oldLExpXBeta = LExpXBeta[k];
        int Lk = L[k];        
        
        // Compute new values
        real newXBeta = oldXBeta + DeltaBeta;
        real newLExpXBeta = Lk * exp(newXBeta);                
        
        // Write sparse elements, many non-coalesced
        XBeta[k] = newXBeta;
        LExpXBeta[k] = newLExpXBeta;        
        
        // Enforce single thread access at any time
        atomicAdd(&NLExpXBeta[n], (newLExpXBeta - oldLExpXBeta));        				
    }					
}
\end{lstlisting}
\renewcommand{\baselinestretch}{\textSpacing} 


\paragraph{Precision}

Graphics rendering traditionally requires at most 32-bit (single precision) floating-point computation to encompass 8-bits of red, green, blue and alpha.
Ensuingly, GPU performance remains greatest at single precision.
While the latest generations of GPUs can operate with 64-bit (double precision) numbers, the precision boost comes with a performance cost because the GPU contains fewer double precision arthimetic logical units, resulting in approximately half the maximum floating-point operations per second.  
Further, double precision mandates reading and writing twice as much information.
For fitting the BSCCS model to massive datasets, single precision arthimetic suffices; the computations do not involve substracting approximately equal quantities, nor multiplying small quantities, both of which may lead to underflow.
To demonstrate this point, finding $\betamap$ for BSCCS is a convex optimization problem \citep{simpson2011self}.  
Subsequently, $\hessianLogLike < 0$ and all elements of $\left[ \mess \times \left(\bf{1} - \mess \right) \right] \ge 0$.  
Likewise, all elements of $\alloffs$, $\exp\left( \XBeta \right)$ and, therefore, $\mess \ge 0$.
\par


\subsection{Hyperparameter Selection and Measures of Coefficient Uncertainty}

\newcommand{\fold}{10}

\myedit{Limits}{
We aim to provide a full Bayesian analysis of all unknown parameters in our model.  However, at present this remains beyond our computational limits.  As a stopgap solution, we borrow two frequentist Monte Carlo proceedures.
}
We learn about the hyperparameter $\sigma^2$ 
through a $\fold$-fold cross-validation scheme. 
Previously, the computational cost of fitting the BSCCS model was too great to consider the hyperparameter as random, requiring an arbitrarily fixed value.
Under this cross-validation scheme, we randomly separate the cases-only dataset into $\fold$ portions, fit the BSCCS model via CCD on $\fold - 1$ of these portions and compute the predictive log-likelihood $\logLike \left( \bs{\beta} \right)$ of the remaining portion given the fit.
We repeat this process across a log-scale grid of hyperparameter values and chose the hyperparameter that maximizes the predictive log-likelihood.
To moderately reduce the number of iterations required to acheive convergence of the CCD algorithm for successive hyperparameter values, we order the grid values from smallest to largest prior variance and exploit a series of ``warm-starts.''  
At small variance under the Laplacian prior, most coefficients shrink to $0$ and only slowly enter into the regression as the variance increases \citep{wu2009genome}.  Under the warm-start, the maximized regression coefficients from the previous fit serve as starting values for the next fit.
In general, the predictive log-likelihood surface is relatively flat in the region around its maximum, 
so precise estimation of the hyperparameter is unnecessary.
Alternative maximization strategies involving an initial bracketing of the maximized predictive log-likelihood and an intervaled line search often yield more precise estimates in fewer evaluations of the predictive log-likelihood.

\newcommand{\bsfold}{200}
\par
Along with the infeasibility of estimating the hyperparameter, generating measures of uncertainty on the regression coefficients has remained taxing, to say the least, for the BSCCS model applied to massive observational databases.
As a first attack at this problem, we examine the non-parametric bootstrap \citep{efron1986bootstrap} in the context of a $L_1$ regularized GLM \citep{park2007l1}.
Procedures for generating standard errors for parameter
estimates in the context of $L_1$ or $L_2$ regularization
that are both computationally efficient and theoretically
well-supported remain out of reach. The simple non-parametric
bootstrap approach we pursue here has some short-comings
(see \citet{chatterjee2010bootstrapping} for a related discussion
in the context of linear regression), but we view it as a
pragmatic approach pending a more complete solution.
Without getting embroiled in this discussion, we report 95\% confidence intervals derived from the 2.5\% and 97.5\% quantiles of $\bsfold$ bootstrap samples.
Under the Laplacian prior and as a ``poor man's estimate'' of the posterior probability that $\beta_{\drug} \neq 0$, we also report $\hat{p}_{\drug}$ for each drug, the observed bootstrap proportion in which $\betamapj{\drug}$ achieves a non-zero MAP estimate.

\section{Demonstration}

We examine the computational performance of fitting the BSCCS model across several large-scale observational databases and AEs.
In particular, we show results from two medical claims databases and 
acute liver injury, acute renal failure, bleeding and upper gestrointestinal tract ucler hospitalization events
in order to provide an examplar range of dataset sizes.
\myedit{Disclaimer}{
Our results explore the effects of optimization and parallelization and are not meant here to identify medical products associated with these events.
}
The MarketScan\texttrademark\ Commercial Claims and Encounters (CCAE) Research Database from Thomson Reuters is a large administrative claims database containing 59 million privately insured lives and provides patient-level deidentified data from inpatient and outpatient visits and pharmacy claims of multiple large employers.  
The MarketScan Lab Database (MSLR) contains 1.5 million persons representing a largely privately-insured population, with administrative claims from inpatient, outpatient, and pharmacy services supplemented by laboratory results.
These databases constitute part of the data community established within the 
the Observational Medical Outcomes Partnership (OMOP).  
OMOP is a public-private partnership between government, industry and academia to conduct methodological research to inform the appropriate use of observational healthcare data for active medical product surveillance.

\par

These example datasets span $N = 115\mbox{K}$ to $3.6\mbox{M}$ cases-only patients taking $\numDrugs = 1224$ to $1428$ different drugs. The datasets provide $K = 3.8\mbox{M}$ to $75\mbox{M}$ total (unique) exposure eras per analysis. 
We perform all benchmarking on the Amazon Elastic Compute Cloud, exploiting an Intel Xeon X5570 CPU @ 2.93GHz and one NVIDIA Tesla C2050. 
This GPU device sports 448 cores @ 1.15GHz.
\myedit{Amazon}{
Performance on less expensive, commodity-grade GPUs, such as the NVIDIA GTX580, is often greater due to a slightly larger number of cores per GPU and higher memory-bandwidth.  Due to data licensing agreements, however, we are restricted to Amazon hardware.
}

\par
\begin{figure}[htb]
\centerline{\includegraphics[width=0.5\textwidth]{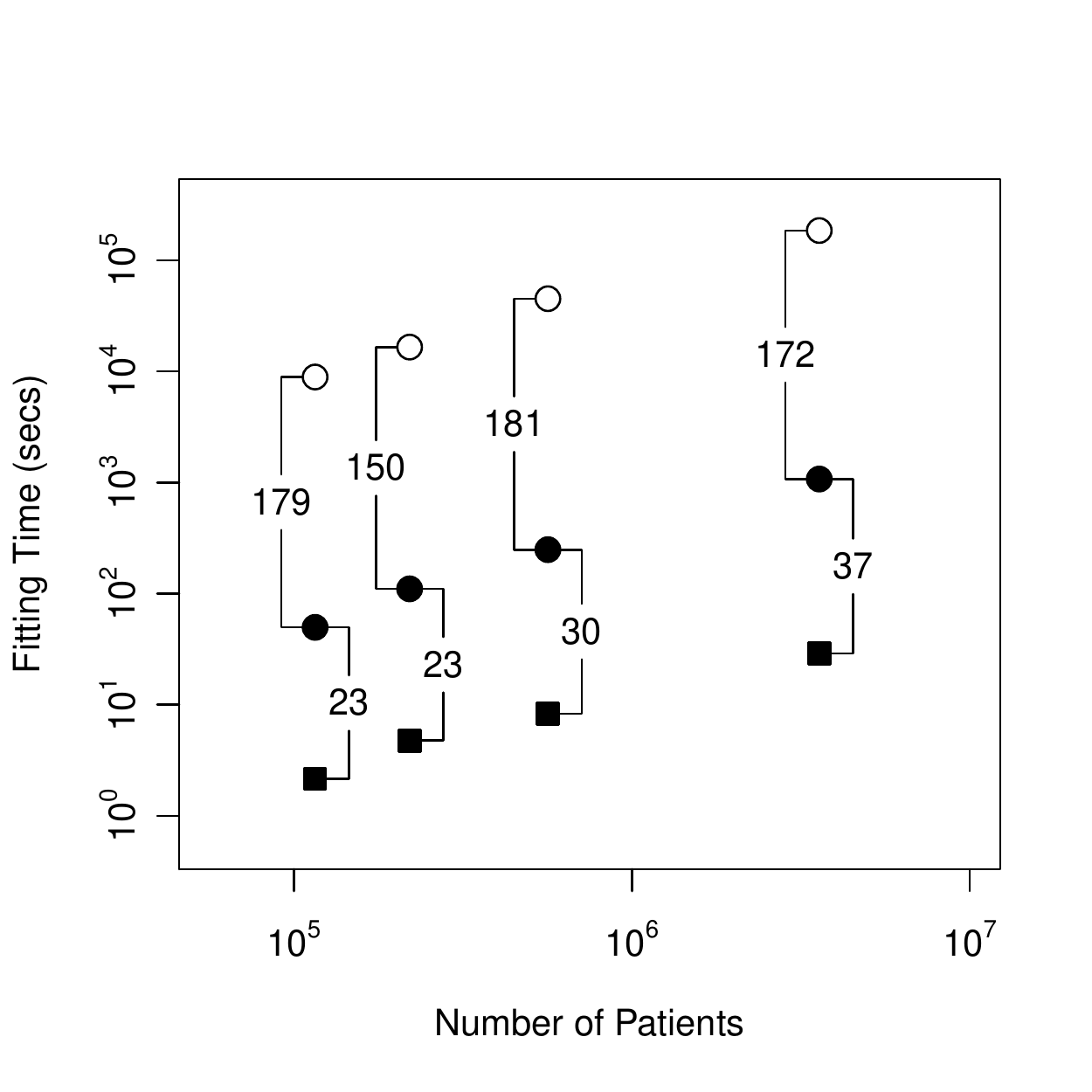}}
\caption{Maximum \textit{a posteriori} estimation for several observational databases under the Bayesian self-controlled cases series model.
We provide run-times for three implementations: dense computation on the CPU (white circles), sparse computation on the CPU (black circles) and sparse computation on the GPU (black squares).
}
\label{speed-up-figure}
\end{figure}
Figure \ref{speed-up-figure} presents the relative speed-up our algorithms enjoy when inferring MAP estimates.
These gains first compare implementing Steps \ref{exp-step} - \ref{numerator-step} as sparse operations and then porting computing to the GPU.
Sparsity generates up to a 181-fold speed-up; while the GPU multiplies this by up to another 37-fold.
To put these times on an absolute scale, MAP estimation for our largest dataset originally drained over 51 hours; sparse operations on the GPU reduce this time to 29 seconds.
Naturally, with fitting times standing in the 10s of hours, the hopes for cross-validation or bootstrap remain low, but grow very practical at 10s of seconds per replicate.


\par
\begin{figure}[htb]
\centerline{\includegraphics[width=0.75\textwidth]{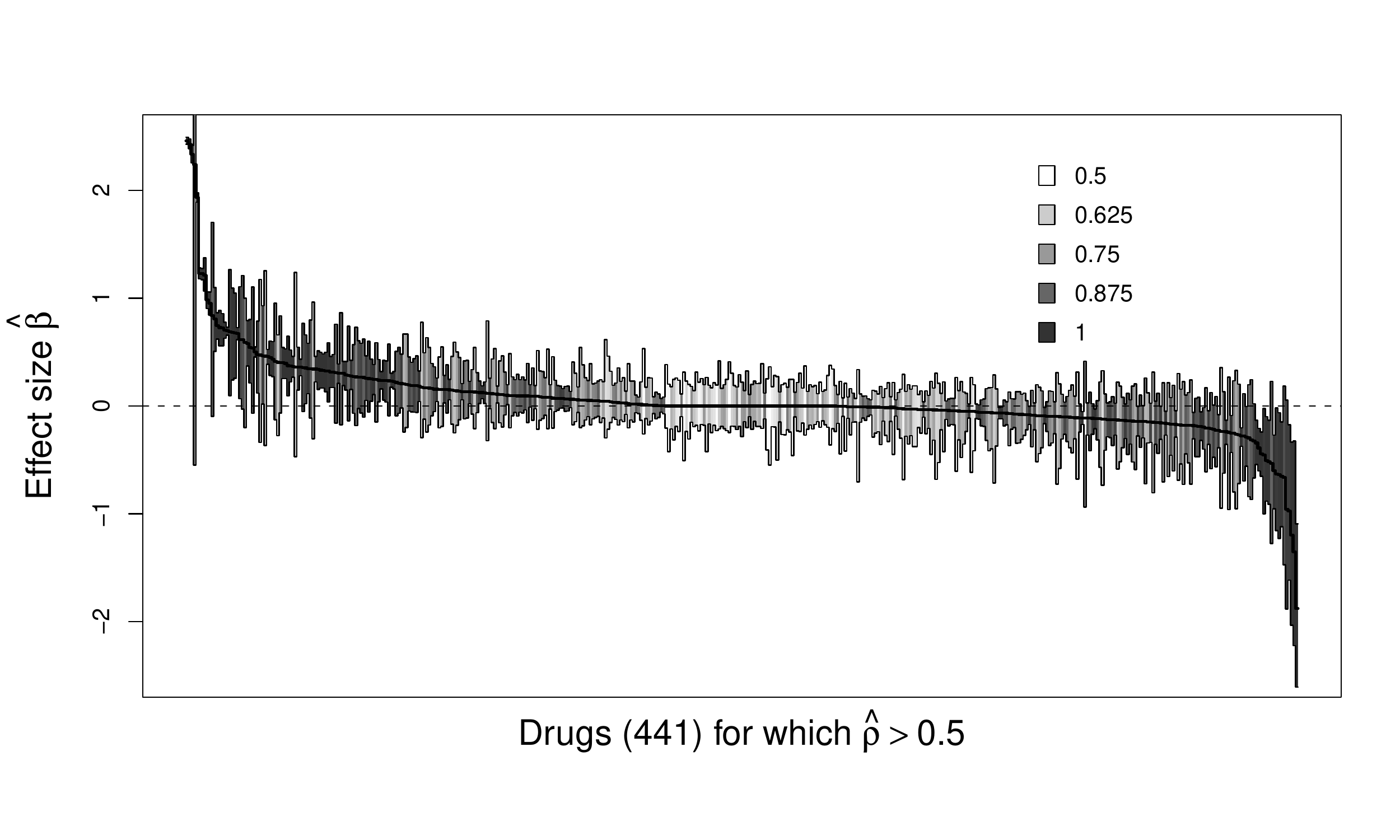}}
\caption{Exemplar uncertainty analysis of angioedema as an adverse event under the $L_1$ prior.  
Here, we plot the non-parametric bootstrap 95\% confidence intervals for the $441$ drug effects that demonstrated non-zero coefficients in at least 50\% of the bootstrap replicates. Gray-scaling reports the proportion of bootstrap replicates in which effect estimates are non-zero.
}
\label{boot-strap-figure}
\end{figure}
Cross-validation to learn the hyperparameter $\sigma^2$ across these four datasets returns optimal variances ranging from $0.05$ to $0.15$ for $L_1$ and $0.02$ to $0.13$ for $L_2$.
%
%
%
%
%
%
Importantly, these ranges are approximately an order-of-magnitude smaller than the arbitrary fixed value previously assumed in our BSCCS studies for drug surveillance.
Employing the optimal hyperparameter, Figure \ref{boot-strap-figure} reports non-parameteric bootstrap confidence intervals of drug effects for a single representative dataset under the $L_1$ prior.
This dataset explores angioedema events within the CCAE database and contains $N = 76\mbox{K}$ case-only patients, taking $\numDrugs = 1162$ drugs and yielding $K = 2.1\mbox{M}$ exposure eras.
In the figure, we first rank all drugs by their MAP estimate $\hat{\beta}_j$ in decreasing order and then plot the 95\% confidence intervals for the 441 drugs for which $\hat{p}_{\drug} > 0.50$.
Darker interval shading reflects larger $\hat{p}_{\drug}$.
While a general trend holds in which larger $| \hat{\beta}_j |$ more often return 95\% confidence intervals that do not cover $0$, we identify notable exceptions.   
Namely, Drotrecogin alfa, an anti-thrombotic, anti-inflammatory agent used in the treatment of severe sepsis, returns with the fourth largest effect estimate, but its confidence interval continues to cover $0$, reflecting the high sampling variability in this estimate.

%

\section{Discussion}

Efficient algorithmic design and massive parallelization open the door for fitting complex GLMs to massive datasets.  
Computational statisticians regularly capitalize on the sparsity of their model and data, and this is an important design issue for the BSCCS we consider here, since the design matrix ${\bf X}$ consists purely of sparse covariates.
In particular, we identify that the sparsity of ${\bf X}$ carries all the way through to computing the subject-specific contributions to the model gradient and Hessian, resulting in over a $100$-fold speed-up compared to the most advanced CCD algorithms for GLMs of which we are aware.
Many GLMs, however, command dense covariates as well, such as baseline measurements, and other techniques become necessary.
Here fusing multiple transformations and reduction together into vectorizable kernels is the first step in off-loading the work to the GPU, and we hope our discussion in this paper raises awareness of these techniques among computational statisticians.
The end result for the BSCCS model is an approximate $30$-fold speed-up on a single GPU compared to a single CPU core.
These techniques also port directly to utilizing multi-core CPUs and multiple GPUs simultaneously, although we do not explore this avenue in this paper to simplify comparisons.

Advancing model complexity is both possible in the sampling density of the data and in the prior assumptions on the unknown model parameter.
Here we have only considered independent and identically distributed prior densities over the drug effect sizes.
More biologically plausible hierarchical distributions are conveniently available.  
For example, to borrow strength, we may favor grouping drugs \textit{a priori} into classes based on mode of action or therapeutic targets.
Similarly, we may wish to explore borrowing strength across related outcomes.
Because computation of the prior gradient and Hessian remains extremely light-weight, no modification to the GPU code is necessary and run-times should remain as quick.

\par
\myedit{RewriteSentence}{
One immediate advantage of the orders-of-magnitude reduction in run-time stands the ability to nest point-estimation within both cross-validation and bootstrap frameworks, making these Monte Carlo frameworks feasible.
Cross-validation and bootstrapping begin to allow us to estimate model hyperparameters and report measures of uncertainty around the usual point-estimates.
}
For the drug surveillance community, this represents a giant leap forward.
For example, most of the statistial methods in OMOP are implemented in the statistical packages {\sf SAS} or {\sf R} (\url{http://omop.fnih.org/MethodsLibrary}).
\myedit{NaiveTools}{
One of our own preliminary implementations of the BSCCS model in {\sf R}, using just a sparse matrix package and no further linear algebra libraries, requires around 5.3 hours to generate a single MAP estimate from a dataset with only $N = 7460$.}
With this benchmark in mind, it is no wonder why almost all computationally expensive fitting of massive datasets in the field has ignored cross-validation and bootstrapping; see, e.g., \citet{funk2011doubly}, and
a presentation at the 2011 International Congress on Pharmacoepidemiology involving a similar study redoubled this point by claiming that bootstrapping is computationally infeasible with more than $20\mbox{K}$ patients.
High performance statistical computing involving massive parallelization shows that these limitations are quickly lifting.

\par

We achieve this success by
exploiting the GPU within a serial CCD algorithm. 
CCD is a generic optimization approach and we envision extensions working for massive dataset applied to models beyond the GLM setting as well.
\myedit{OtherUses}{
Moving past CCD, \cite{zhou2010graphics} consider similar block-relaxation and majorization techniques to attack large-scale matrix factorization and multidimensional scaling using GPUs.  Here and in CCD, one breaks a high-dimensional optimization problem into a series of low-dimensional updates that involve many scalar operations.  As \cite{zhou2010graphics} demonstrate with their quasi-Newton acceleration application, 
one ideally aims for one-dimensional updates, as even slightly higher-dimensional operations carry heavy data-dependency that can outweigh the advantages of the GPU. 
}
\par
To accomplish parallelization within a serial algorithm, we take advantage of the wide vector-processing capabilities of the GPU to perform simple operations simultaneously across a large input of data.  This vectorization lacks branches in the kernel code, avoiding thread divergence and serialization of the work within the wide vector.
As a result, 
expected speed-up
scales most directly with the quantity of data. 
\myedit{NotEP}{
This differs considerably from distributing EP tasks, such as those that arise in many Monte Carlo approaches including the independent and often divergent particle evolution in a sequential Monte Carlo, to separate cores of the GPU.  
}
Here, we receive at little cost more particles and higher precision estimates with additional cores.
Unfortunately, however, this strategy loses out on scaling in the critical dimension of the data as massively parallel devices continue to mushroom in size.

\section{Acknowledgments}

The Observational Medical Outcomes Partnership is funded by the Foundation for the National Institutes of Health through generous contributions from the following: Abbott, Amgen Inc., AstraZeneca, Bayer Healthcare Pharmaceuticals, Inc., Bristol-Myers Squibb, Eli Lilly \& Company, GlaxoSmithKline, Johnson \& Johnson, Lundbeck, Inc., Merck \& Co., Inc., Novartis Pharmaceuticals Corporation, Pfizer Inc, Pharmaceutical Research Manufacturers of America (PhRMA), Roche, Sanofi-Aventis, Schering-Plough Corporation, and Takeda.
MAS is funded in part by the National Institutes of Health (R01 HG006139) and a research award from Google. 


\bibliography{bsccs}

\end{document}